\documentclass[twocolumn,aps,pre,superscriptaddress]{revtex4-1}
\usepackage[utf8]{inputenc}
\usepackage{epsf,graphicx}
\usepackage{multirow}
\usepackage{amsmath,gensymb,amssymb}
\usepackage{siunitx}
\usepackage[bottom]{footmisc}
\usepackage{lipsum}
\usepackage{dcolumn}
\usepackage{bm}
\usepackage{hyperref}
\hypersetup{
colorlinks=true, 
citecolor=magenta,
breaklinks=true, 
urlcolor= blue, 
linkcolor= blue, 
bookmarksopen=true, 
}

\usepackage{natbib}

\usepackage{color}
\definecolor{linkcolor}{rgb}{0,0,0.6}

\begin{document}

\title{Three-dimensional direct numerical simulation of free-surface\\ magnetohydrodynamic wave turbulence}	

\author{Evgeny Kochurin}
\email{kochurin@iep.uran.ru}
\affiliation{Skolkovo Institute of Science and Technology, 121205, Moscow, Russia}
\affiliation{Institute of Electrophysics, Ural Division, Russian Academy of Sciences, 620016, Yekaterinburg, Russia}

\author{Guillaume Ricard}
\affiliation{Universit\'e Paris Cité, CNRS, MSC Laboratory, UMR 7057, F-75013 Paris, France}

\author{Nikolay Zubarev}
\affiliation{Institute of Electrophysics, Ural Division, Russian Academy of Sciences, 620016, Yekaterinburg, Russia}
\affiliation{Lebedev Physical Institute, Russian Academy of Sciences, 119991, Moscow, Russia}

\author{Eric Falcon}
\email{eric.falcon@u-paris.fr}
\affiliation{Universit\'e Paris Cité, CNRS, MSC Laboratory, UMR 7057, F-75013 Paris, France}
\date{\today}

\begin{abstract}
We report on three-dimensional direct numerical simulation of wave turbulence on the free surface of a magnetic fluid subjected to an external horizontal magnetic field. A transition from capillary-wave turbulence to anisotropic magneto-capillary wave turbulence is observed for an increasing field. At high enough field, wave turbulence becomes highly anisotropic, cascading mainly perpendicularly to the field direction, in good agreement with the prediction of a phenomenological model, and with anisotropic Alfv\'en wave turbulence. Although surface waves on a magnetic fluid are different from Alfv\'en waves in plasma, a strong analogy is found with similar wave spectrum scalings and similar magnetic-field dependent dispersionless wave velocities.
\end{abstract}

\maketitle
\textit{Introduction.}---
Most of nonlinear wave systems reach a wave turbulence regime as a result of wave interactions~\cite{ZakharovBook,NazarenkoBook}. This phenomenon occurs in various domains at different scales such as ocean surface waves, plasma waves, hydroelastic or elastic waves, internal or inertial waves, and optical waves \cite{NazarenkoBook}. The weakly nonlinear theory (called weak turbulence theory) derived analytically the solutions of the corresponding kinetic equations \cite{ZakharovBook,NazarenkoBook,ZakharovSPD67grav,ZakharovJAMTP67cap,Zakharov1970}. These solutions, known as the Kolmogorov-Zakharov (KZ) spectra, describe the energy transfers towards small scales (direct cascade) or large ones (inverse cascade). Athough these solutions have been tested in different systems, numerical and experimental works are currently a paramount of interest to understand in what extend this theory can describe real physical systems.

One of the most important system is Alfv\'en waves in magnetohydrodynamics (MHD) \cite{Alfven1942}, initially observed in laboratory plasma \cite{Lundquist1949,BostickPR52,LehnertPR54,AllenPRL59}, and recently in astrophysical plasma such as the Sun's outer \cite{DePontieu2007} or inner \cite{KasperPRL2021} atmosphere. Three-dimensional (3D) Alfv\'en waves in a turbulent regime were initially predicted to follow the isotropic Iroshnikov-Kraichnan spectrum \cite{Iroshnikov1963,KraichnanPoF1965}. However, they become strongly anisotropic in a presence of an intense magnetic field, and transfer energy mainly in the plane transverse to the field, thus becoming nearly two-dimensional~\cite{GaltierBook,ShebalinJPP1983,NazarenkoBook}. The spectrum of this anisotropic weak turbulence regime has been derived \cite{NgPP1997,GaltierJPP2000}, then observed in the Jupiter's magnetosphere~\cite{Saur2002}, and confirmed recently numerically \cite{Meyrand2015,Meyrand2016}. An analogous anisotropic behavior is predicted for hydrodynamics waves on the surface a magnetic fluid subjected to a horizontal magnetic field~\cite{RosensweigBook}. Although wave turbulence regimes have been observed on the surface of a ferrofluid in an external magnetic field both experimentally~\cite{BoyerPRL2008,DorboloPRE2011} and numerically \cite{KochurinJMMM20}, the anisotropic regime has never been reported so far in such a system, to our knowledge.

In this Letter, we show the existence of an analogy between Alfv\'en wave turbulence and wave turbulence on the surface of a magnetic fluid. The analogy is not only qualitative but also quantitative in term of cascade of energy. In particular, we show that an anisotropic MHD wave turbulence emerges at high enough magnetic field with a wave spectrum showing similar scalings than the ones of anisotropic Alfv\'en wave turbulence predictions. 

\textit{Theoretical backgrounds.}--- We consider an ideal incompressible magnetic liquid of infinite depth subjected to an external horizontal magnetic field $B$ directed along the $x$-axis. The dispersion relation of linear waves on the surface of such a ferrofluid reads, neglecting gravity, ~\cite{melcher1961}
\begin{equation}
\label{disp}
\omega^2(\bm{k})=[B^2/(\tilde{\mu}\rho)]k_x^2+(\gamma/\rho)k^3 {\rm \ ,}
\end{equation}
where $k\equiv |\bm{k}|=\sqrt{k_x^2+k_y^2}$ is the wave number, $\omega$ is the angular frequency, $\gamma$ and $\rho$ are the surface tension and mass density of the liquid, $\tilde{\mu}=\mu_0(\mu+1)/(\mu-1)^2$, $\mu_0$ is the magnetic permeability of vacuum, and $\mu$ is the relative permeability of the liquid. Equation~\eqref{disp} describes the anisotropic propagation of surface waves. We define $v_A^2=B^2/(\tilde{\mu}\rho)$, the group velocity of dispersionless magnetic surface wave propagating along $B$ (analogous to Alfv\'en velocity \cite{Alfven1942}). Note that $\mu$ is here constant, whereas $\mu$ experimentally depends of $B$ \cite{DorboloPRE2011}. This change has no impact here since $v_A$ is the parameter used to quantify magnetic effects.

The power spectrum of wave elevation $S(k)$ is defined as the square modulus of the Fourier transform of the wave elevation $\eta(x,y)$. Without magnetic field, the KZ spectrum for isotropic capillary wave turbulence reads~\cite{ZakharovJAMTP67cap}
\begin{equation} \label{KZ}
S^c(k)=C_{KZ}P^{1/2}\left(\gamma/\rho\right)^{-3/4}k^{-15/4} {\rm \ ,}
\end{equation}
where $C_{KZ}$ is the nondimensional KZ constant and $P$ is the energy flux per unit area and density. 
The energy spectrum is $E^c(k)=(\gamma/\rho)k^2S^c(k)$. To date, the KZ spectrum has been very well confirmed for capillary waves both experimentally (e.g., see \cite{FalconPRL07,FalconEPL09,KolmakovPLTP09,ARFM2022}) and numerically \cite{PushkarevPRL96,PushkarevPD00,DeikePRL14,PanPRL14,PanJFM15} for weakly nonlinear capillary waves. For $B\neq0$, no weak turbulence prediction for magneto-capillary waves exists so far, only dimensional analysis has been done~\cite{BoyerPRL2008,DorboloPRE2011}. For Alfv\'en waves in a plasma within a strong magnetic field (e.g., along $x$), the energy transfer by three-wave interactions has been shown to be frozen in the field direction and to occur only in the transverse direction to $B$ \cite{ShebalinJPP1983}. The weak turbulence predictions for the power spectrum of such anisotropic Alfv\'en wave turbulence ($k_y\gg k_x$, i.e., $|\bm{k}|\sim k_y$) reads \cite{GaltierJPP2000,NgPP1997,GaltierBook}
\begin{equation} \label{AlfvenSpectrum}
S^{m}(k)=C_{m}P^{1/2}v_A^{-3/2}k_y^{-3},
\end{equation}
where $C_{m}=1.467$~\cite{GaltierBook2} (a $\sqrt{2\pi}$ factor was missing in~\cite{GaltierJPP2000}). The energy spectrum is $E^m(k)=v_a^2kS^m(k)$.

\textit{Model equations.}--- The numerical model used here is based on the Hamiltonian equations describing the MHD motion of an ideal irrotational and incompressible ferrofluid subjected to an external horizontal magnetic field. We assume the absence of free electric charge and current in the fluid, which means that the magnetic field in the liquid is also potential. In the quadratic nonlinear approximation, the equations of boundary motion are written as
\begin{equation}\label{eq1}
\eta_t=\hat k \psi-\hat k(\eta \hat k \psi)-\nabla_{\perp}(\eta \nabla_{\perp}\psi)+\hat D_k\eta {\rm \ ,}
\end{equation}
\begin{multline}
\psi_t=\nabla_{\perp}^2\eta+\frac{1}{2}\left[(\hat k \psi)^2-(\nabla_{\perp}\psi)^2\right]+V_A^2\hat k^{-1}\eta_{xx} \\
\left. -\frac{A_{\mu}V_A^2}{2}\left[2\hat k^{-1} \partial_x\left(\eta \hat k \eta_x-\nabla_{\perp} \eta \cdot \nabla_{\perp}\hat k^{-1}\eta_x \right)-\eta_x^2\right.\right.\\
\left.-2\eta\eta_{xx}-(\nabla_{\perp}\hat k^{-1}\eta_x)^2\right]+\mathcal{F}(\bm k,t)+\hat D_k \psi {\rm \ ,}
\label{eq2}
\end{multline}
where $\nabla_{\perp}=\{\partial_x, \partial_y\}$ is the nabla operator, $\psi$ the velocity potential, $\hat k$ is the integral operator having the form $\hat k f_k=k f_k$, $\hat k ^{-1}$ is the inverse $\hat k$-operator, $V_A^2=v_A^2[\rho/(g\gamma)]^{1/2} $ is the nondimensional MHD wave speed, $g$ is the gravity acceleration, and $A_{\mu}=(\mu-1)/(\mu+1)$ is the magnetic Atwood number. $\hat D_k$ is the viscosity operator acting as, $\hat D_k f_k=-\nu (k-k_d)^2 f_k$, for $k\geq k_d$, and, $\hat D_k=0$, for $k< k_d$, the coefficient $\nu$ determines the intensity of energy dissipation (see \cite{zu04,zu09,zuko1}). More details on the derivation of Eqs.~\eqref{eq1} and ~\eqref{eq2} from potential equations are given in the Supplemental Material~\cite{SuppMat}. The pumping term $\mathcal{F}(\bm{k},t)$ in Eq.~\eqref{eq2} is defined in Fourier space as $\mathcal{F}(\bm{k},t)=F(k) \exp[i\omega(\bm{k})t]$, where $ F(k)=F_0\exp[-(k-k_0)^4/k_f] $, with $F_0$ is the forcing amplitude reached at $k=k_0$. The wave vectors are pumped in the Fourier space in the range $k\in[1, k_f]$ (see below), and in all directions. In the absence of dissipation and pumping, exact analytical solutions of Eqs.~\eqref{eq1}-\eqref{eq2} has been found in the strong-field limit and $\mu \gg 1$ \cite{zu04,zu09}.

For finite $\mu$, the surface waves collapse under the action of infinitely strong horizontal field \cite{kozuzu}. Thus, for a correct simulation of the free surface MHD wave turbulence, it is necessary to take into account the regularizing effects of viscosity and surface tension.

\begin{figure}[t!]
	\centering
	\includegraphics[width=1.0\linewidth]{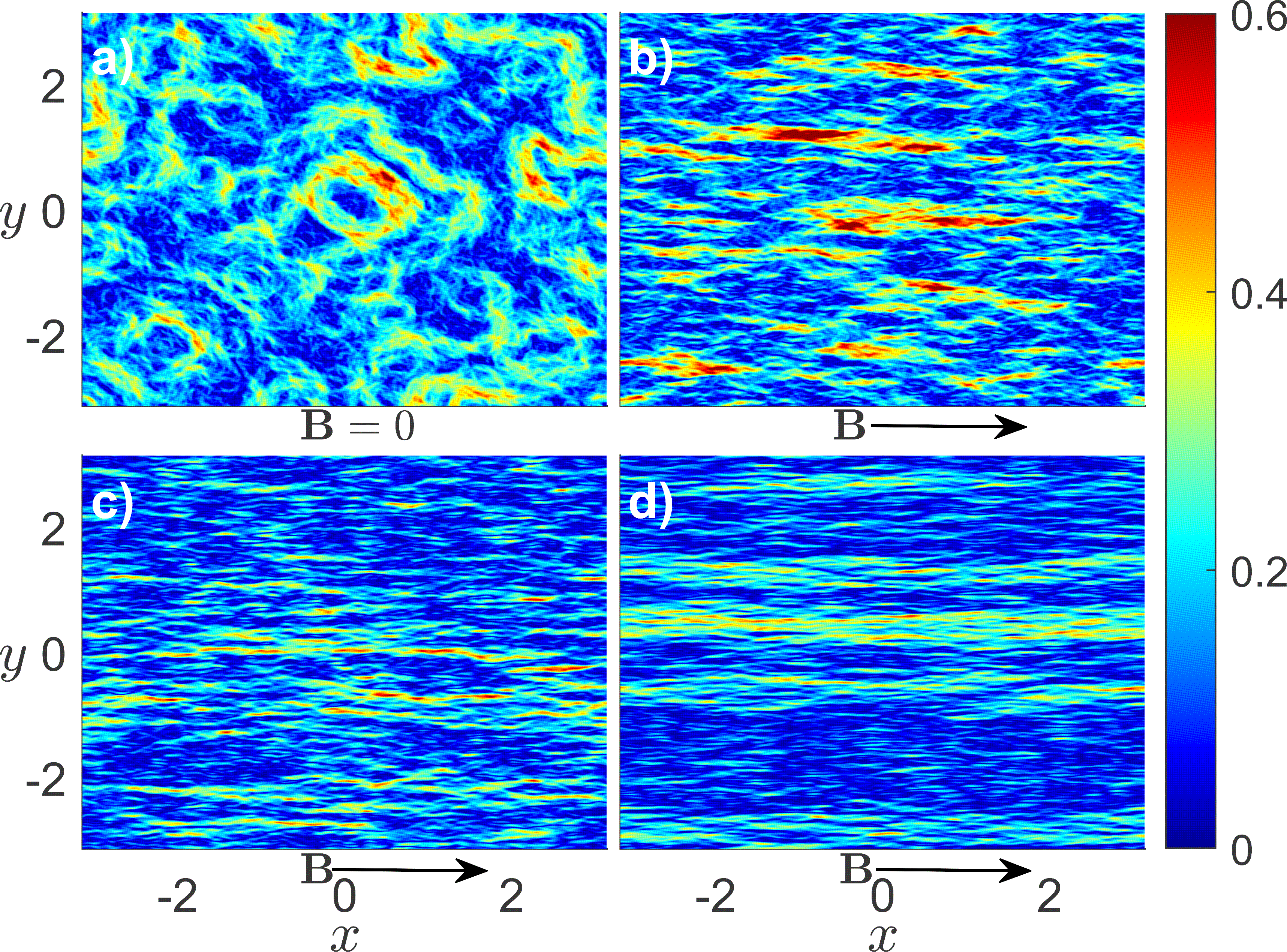}
	\caption{Free surface gradient at a fixed time in the steady state ($t=250$) for different values of $B$: $V_A^2=$ (a) 0, (b) $25$, (c) $100$, and (d) $300$. $B$ is along the $x$-axis.}
	\label{fig1}
\end{figure}

\begin{figure*}[t!]
    \includegraphics[width=1.0\linewidth]{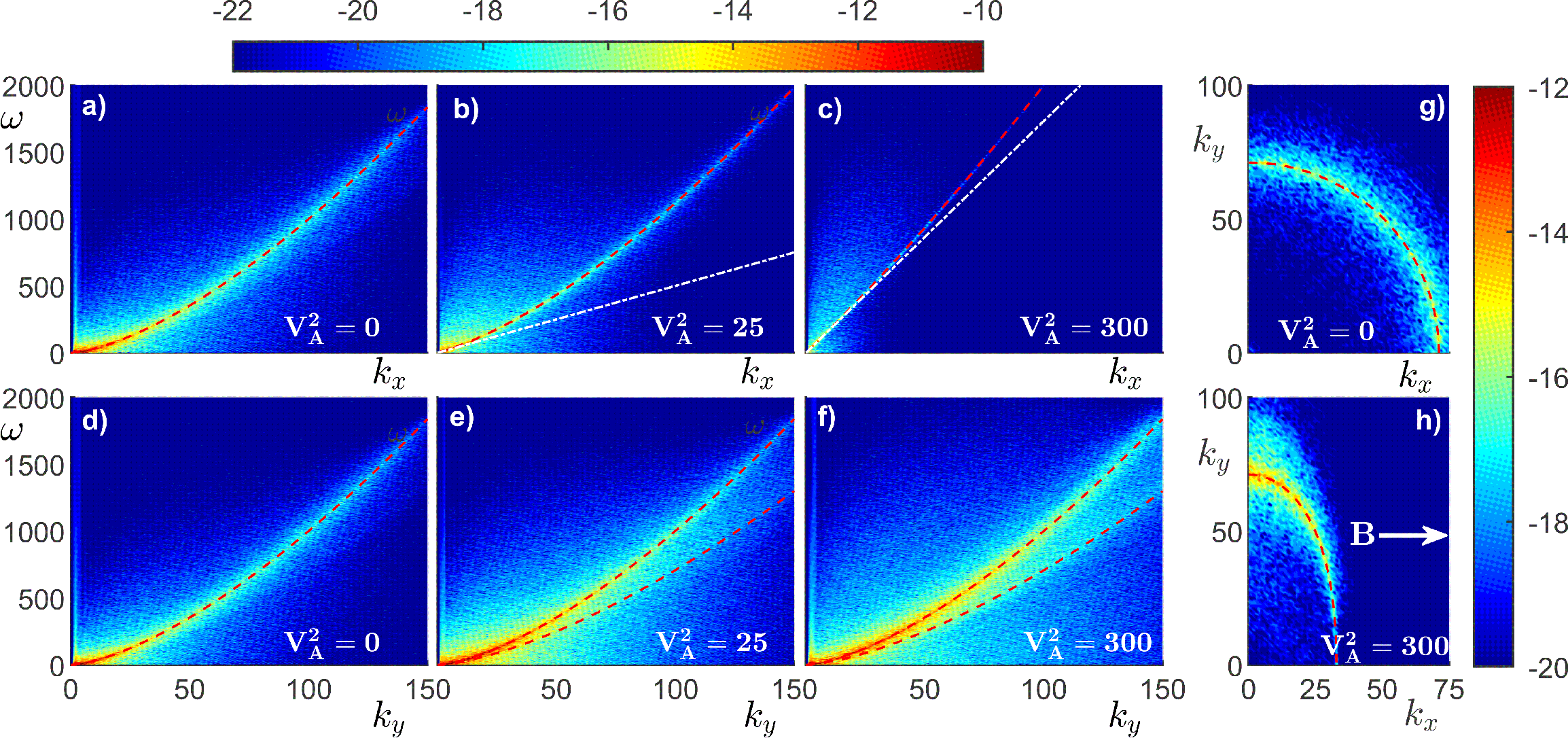}
	\caption{(a)-(c) Power spectra $S(k_x,\omega)$ of surface waves in the field direction ($x$-axis) for different $B$, i.e., different $V_A$. Log-color bar. The red-dashed lines correspond to Eq.~\eqref{disp} (dispersion relation) and white dash-dotted lines to nondispersive wave propagation, $\omega=V_A k_x$. (d)-(f) Power spectra $S(k_y,\omega)$ of waves traveling along $y$-axis. The red dashed lines correspond to $\omega=k_y^{3/2}$ and $\omega=k_y^{3/2}/2^{1/2}$. (g) and (h): Cross sections $S(k_x,k_y,\omega^{\star})$ of the power spectrum at a fixed frequency $\omega^{\star}=600$ for different fields $V_A^2=0$ and $V_A^2=300$, respectively. Red-dashed lines correspond to $|\bm{k}(\omega^{\star})|$ using Eq. \eqref{disp}.}
	\label{fig2}
\end{figure*}

Equations~\eqref{eq1} and \eqref{eq2} are solved numerically using the pseudo-spectral methods with the total number of Fourier harmonics $N\times N$. The time integration scheme is based on the explicit fourth-order Runge-Kutta method with a step $dt$. To stabilize the numerical scheme, a low-pass anti-aliasing filter is used~\cite{PushkarevPRL96}. At each integration time step, harmonics with wavenumbers greater than $k_a$ are equated to zero. The effect of this low-pass filtering can be thus interpreted as ``superviscosity'' acting at small scales. Simulations are performed in a periodic box of size $2 \pi \times 2 \pi$ with $N=1024$, $dt=5\times 10^{-5}$, $F_0=2000$, $k_0=3$, $k_f=6$, $k_d=150$, $k_a=212$, and $\nu=10$. All numerical simulations are carried out for a magnetic fluid with $A_{\mu}=0.5$, which corresponds to $\mu=3$. We present below four series of simulations with different values of the magnetic parameter $V_A^2=0$, $25$, $100$, and $300$. The typical wave steepness $\epsilon \equiv \left\langle \sqrt{\int_{\mathcal{S}} ||\nabla \eta(x,y,t)||^2dxdy/\mathcal{S}} \right\rangle_t$ used is $0.16$ and is found to be almost constant when $B$ is increased. The stationary state is reached after a time $t \approx 50$, and each simulation lasts up to $t=500$. 

\textit{Phenomenological analysis}--- The theoretical spectrum expected from our model equations [Eqs.~\eqref{eq1} and~\eqref{eq2}] is obtained following a phenomenological method described in Ref. \cite{GaltierBook,GaltierBook2} and detailed in the Supplemental Material~\cite{SuppMat}. We assume a strong magnetic field ($V_A k_x\gg k_y^{3/2}$) and an anisotropy of the wave field ($k_y\gg k_x$). The fourth member of the rhs of Eq.~\eqref{eq2} provides then an estimation of the nonlinear magnetic timescale as $T_{nl}^m\sim \psi/(V_A^2 k_x^2 \eta^2)$. Assuming that all the magnetic potential energy is transferred to capillary kinetic energy, one has $V_A^2k_x\eta^2\sim\psi^2k_y$, and thus $T_{nl}^m\sim 1/(\psi k_y k_x)$ with $\psi^2\sim k_yE/k_x$. Using the power budget, the energy flux then reads $P\sim k_yE/(\omega T^2_{nl})$, with $E$ the energy spectrum. The power spectrum of wave elevations, $S^m(k)=E(k)/(V_A^2k)$, finally reads $S^m(k)\sim P^{1/2}V_A^{-3/2}k_y^{-3}$, which is found to be the same as the shear-Alfvén wave turbulence prediction of Eq.~\eqref{AlfvenSpectrum}.

\textit{Anisotropic regime.}--- Figure~\ref{fig1} shows the gradient of the free surface, at a fixed time in the steady state, for different $B$. We observe a transition from an isotropic regime [Fig.~\ref{fig1}(a)] to a highly anisotropic regime [Fig.~\ref{fig1}(d)]. Indeed, the surface relief in Fig.~\ref{fig1}(d) becomes almost unidirectional, 
corresponding to surface waves propagating mainly in the direction perpendicular to $B$ (see below). This anisotropy is due to the stabilizing effect of a horizontal magnetic field on a magnetic liquid. Indeed, for waves propagating in the field direction, the field lines pierce the wavy liquid-gas interface, and flatten it in the field direction as a consequence of the field boundary conditions at the interface \cite{RosensweigBook,Walsh2022}. This behavior is close to the anisotropy observed when Alfv\'en waves propagate in the direction of a magnetic field~\cite{NazarenkoBook}, although of a different origin.  

\textit{Nonlinear dispersion relation.}--- The anisotropy is also evidenced by the full power spectrum $S(k,\omega)$ of surface waves. Figures~\ref{fig2}(a)-(c) show the spectrum $S(k_x,\omega)$ for waves traveling along the field direction for different $B$, i.e., different $V_A$. For $B=0$, the energy injected at low $k$ is redistributed within a large range of wave numbers around the linear dispersion relation of Eq.~\eqref{disp}, as expected. When $B$ is increased, the nonlinear dispersion relation is deformed [see Fig.~\ref{fig2}(b)], then becomes quasi-dispersionless in the field direction [see Fig.~\ref{fig2}(c)]. The spectra $S(k_y,\omega)$ of waves traveling normally to the field are shown in Figs.~\ref{fig2}(d)-(f) for different $B$. Figure~\ref{fig2}(d) corresponds to pure capillary waves ($B=0$). When $B$ is increased, waves of higher and higher wavenumbers are generated. Such enhanced energy transfers perpendicular to the field direction when the latter is increased, are a consequence of the anisotropic effect described above. Figures~\ref{fig2}(e) and (f) display also the emergence of a second branch in the dispersion relation. This branch corresponds to bound waves (harmonics due to nonresonant interactions) \cite{HerbertPRL2010}. Figures~\ref{fig2}(g) and (h) show the cross sections $S(k_x,k_y,\omega^{\star})$ of the power spectrum for a fixed frequency value, $\omega^{\star}$. For $B=0$ [see Fig.~\ref{fig2}(g)], the energy is distributed isotropically in all directions along a circle of radius $|\bm{k}(\omega^{\star})|$ (only the first quadrant is shown). When $B$ is increased, the energy is redistributed anisotropically, much stronger in the perpendicular direction than in the field direction [see Fig.~\ref{fig2}(h)]. This effect is reported for all frequencies (see Supplemental Material~\cite{SuppMat}). A stronger nonlinear broadening also appears since bound waves occur normal to the field direction. To sum up, at high magnetic field, anisotropic wave propagation is observed ($k_y\gg k_x$) as expected by weak MHD (or shear-Alfv{\'e}n) wave turbulence as well as appearance of nonlinear coherent structures in the $y$ direction.

\begin{figure}[t!]
	\centering
	\includegraphics[width=1.0\linewidth]{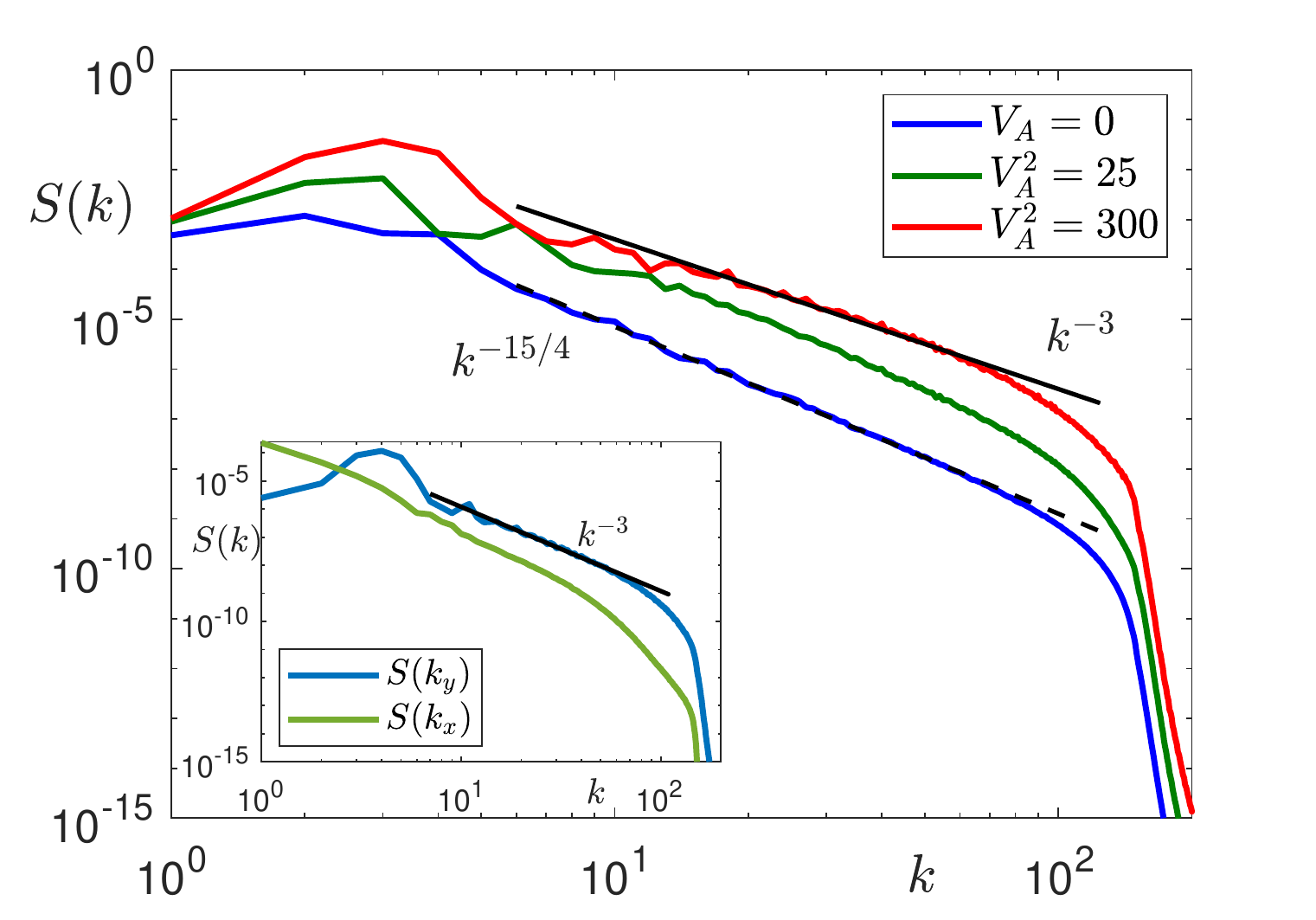}
	\caption{Power spectra $S(k)$ of wave elevations for different $V_A$. Curves have been shifted for clarity. Dashed line: capillary-wave turbulence theory of Eq.~\eqref{KZ} \cite{ZakharovJAMTP67cap}. Solid line: anisotropic MHD wave turbulence theory of Eq.~\eqref{AlfvenSpectrum}. Inset: Spectra in the field direction $S(k_x)$ (green line) and normal to the field $S(k_y)$ (blue line). $V_A^2=300$.}
	\label{fig3}
\end{figure}

\textit{Wave interactions.}--- The MHD surface wave turbulence observed here involved nonlinear waves cascading towards small scales as a result of anisotropic three-wave resonant interactions occurring mainly in the perpendicular direction of the magnetic field (see inset of Fig.~\ref{fig3}). These three-wave resonant interactions are well evidenced by computing the third-order correlation (or bicoherence) of wave elevations (see Supp. Mat.~\cite{SuppMat}).

\begin{figure}[t!]
	\centering
	\includegraphics[width=1.0\linewidth]{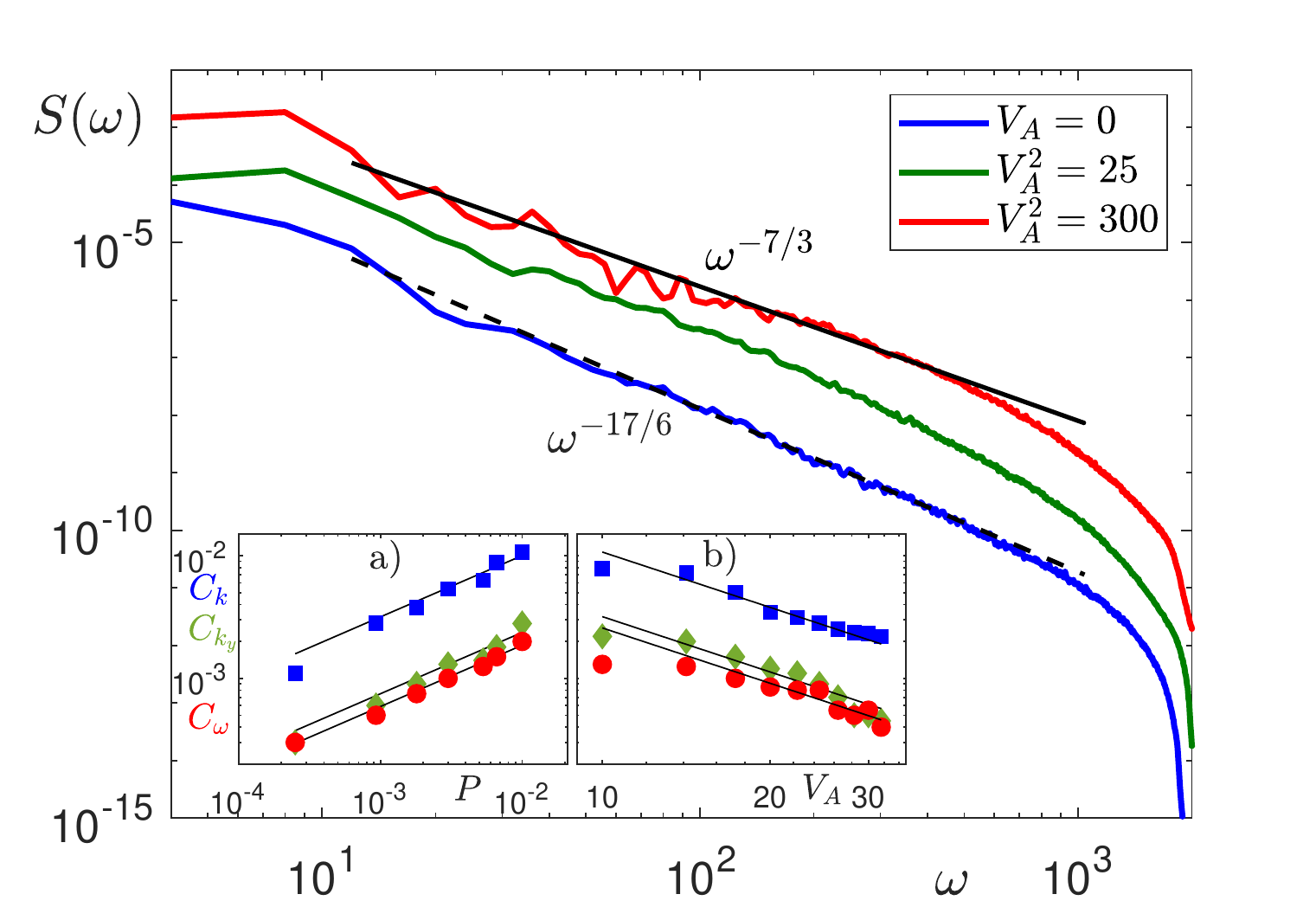}
	\caption{Power spectra $S(\omega)$ of wave elevations for different $V_A$. Curves have been shifted for clarity. Dashed line: capillary-wave turbulence theory in $\omega^{-17/6}$ from Eq.~\eqref{KZ} using $\omega \sim k^{-3/2}$ \cite{ZakharovJAMTP67cap}. Solid line: predictions for anisotropic MHD surface wave turbulence from Eq.~\eqref{AlfvenSpectrum} with $\omega\sim k^{-3/2}$. Inset: spectrum coefficients $C_{k}$ (square), $C_{k_y}$ (diamond) and $C_{\omega}$ (circle) versus (a) $P$ for fixed $V_A^2=300$, and (b) $V_A$ for fixed $P\approx 2.7 \times 10^{-3}$. Solid lines: best fits in $P^{1/2}V_A^{-3/2}$.}
	\label{fig4}
\end{figure}

\textit{Wave-number spectrum.}--- To quantify the transition from isotropic capillary wave turbulence to anisotropic MHD wave turbulence, we compute the power spectra of surface elevations in frequency, $S(\omega)$ and in wave number $S(k)$. $S(k)$ is shown in Fig.~\ref{fig3} for different $V_A$. At zero field, the spectrum is in good agreement with the KZ spectrum in $k^{-15/4}$ of Eq.~\eqref{KZ}. At intermediate field, the spectrum exponent decreases slightly, but is still close to the one predicted by weak turbulence theory. At high enough field, we observe a transition from the KZ spectrum to a spectrum in $k^{-3}$ in rough agreement with the scaling of the phenomenological model and of anisotropic Alfv{\'e}n wave turbulence of Eq.~\eqref{AlfvenSpectrum} (see main Fig.~\ref{fig3}). Specifically, the inset of Fig.~\ref{fig3} shows the spectra $S(k_x)$ along the field direction ($x$-axis) and $S(k_y)$ perpendicular to it ($y$-axis) for $V_A^2=300$. $S(k_y)$ is found to scale as $k_y^{-3}$ as expected by Eq.~\eqref{AlfvenSpectrum} whereas $S(k_x)$ is more than one order of magnitude smaller than $S(k_y)$ in the inertial range. Thus, the spectrum, $S(k)=C_k(P, V_A) k^{-3}$, observed in the main Fig.~\ref{fig3}, is mainly due to the energy transferred perpendicularly to the external field. We determine now the scaling of the coefficient $C_{k}$ with the energy flux $P$ and with the magnetic parameter $V_A$ by performing two series of simulations: (i) at a fixed magnetic field ($V_A^2=300$) for different amplitudes of the energy pumping; (ii) at a fixed rate of energy dissipation $P$ for different fields. The spectrum is then found to increase with the pumping as $C_{k}(P)\sim P^{1/2}$ [see inset (a) of Fig.~\ref{fig4}] and to decrease with the magnetic field as $C_{k}\sim V_A^{-3/2}$ [see inset (b)] for high enough $V_A$. Note that $C_{k_y}$, the coefficient of the spectrum $S(k_y)$, is observed to follow the same scaling. When returning to the dimensional variables, we find thus
\begin{equation}\label{sk}
S(k)\approx S(k_y)=CP^{1/2}v_A^{-3/2}k_y^{-3} {\rm \ ,}
\end{equation}
where $C$ is a constant, independent of $P$ and $V_A$, to be found numerically (see below).
Equation~\eqref{sk} is similar to the spectrum scalings found above by the phenomenological model and to Eq.~\eqref{AlfvenSpectrum} describing anisotropic Alfv\'en waves in plasma subjected to a strong magnetic field. Although MHD surface waves on a magnetic fluid is physically different from MHD Alfv\'en waves in plasma, the anisotropic effect on the energy transfer due to the magnetic field is common and leads to the same scaling for the wave spectrum. Note that, in plasma, no wave propagates normal to the field (only magnetic energy is transferred to the normal direction by shearing) whereas, in our case, capillary waves propagates normal to the field, the magnetic energy being transferred to capillary energy.

\textit{Frequency spectrum.}--- We compute now the frequency spectrum $S(\omega)$ for different $V_A$ as shown in Fig.~\ref{fig4}. At zero field, the simulations again show a good agreement with the predicted KZ spectrum in $S(\omega)\sim \omega^{-17/6}$ \cite{ZakharovJAMTP67cap}. At large enough $B$, the spectrum is less steep and scales as $S(\omega)=C_{\omega} \omega^{-7/3}$. The coefficient $C_{\omega}$ is found to scale as $P^{1/2}$ and $V_A^{-3/2}$  (see insets of Fig.~\ref{fig4}) as for $C_k$. Returning to the dimensional variables thus leads to
\begin{equation}\label{sw}
S(\omega)=C^\prime (\gamma/\rho)^{2/3}P^{1/2}v_A^{-3/2}\omega^{-7/3} {\rm \ .}
\end{equation}
where $C^\prime$ is a constant. Note that the empirical spectra of Eqs.~\eqref{sk}-\eqref{sw} are found to be consistent with each other since they verify $S(\omega)d\omega=S(k)dk \approx S(k_y)dk_y$, using $k_y\gg k_x$ and $\omega(k)$ from Eq.~\eqref{disp}. This also gives $C^\prime=2C/3$.
We find the values of the constants $C=1.7$ and $C^{\prime}=1.34$ (using the best fits of $C_{k_{y}}$ and $C_\omega$ in Fig.~\ref{fig4}a). $C$ is close to the theoretical value $C_m=1.467$ found for shear-Alfv\'en wave turbulence in plasma~\cite{GaltierBook2} and the ratio $2C/3C^\prime=0.85$ is as expected close to 1. Moreover, the spectrum scaling with the energy flux in $P^{1/2}$ is consistent with the fact that three-wave resonant interactions are involved here \cite{NazarenkoBook}. Finally, the timescale separation hypothesis of wave turbulence is verified here since the nonlinear time is found much longer than the linear time regardless of $k$ (see Supplemental Material~\cite{SuppMat}). A typical resonant triad is also shown in [40] to highlight that resonant wave vectors are of the same order of magnitude suggesting that wave interactions are local. Our results thus differ from two-dimensional MHD weak-turbulence predictions leading to nonlocal interactions and a flat steady-state spectrum~\cite{Tronko}. 

\textit{Conclusion.}--- We have numerically reported a transition from isotropic capillary-wave turbulence to a strongly anisotropic MHD wave turbulence on a surface of a magnetic fluid, for high enough magnetic field. In this anisotropic regime, the wave spectrum is found to be in good agreement with the prediction of three-dimensional weak MHD (shear-Alfv\'en) wave turbulence. This highlights the broad application of weak wave turbulence where different physical systems can lead to similar phenomena, only because they share similar dispersion relations and the same nonlinear interaction process (three-wave resonant interactions). Such quantitative analogy between Alfv\'en waves in plasma and surface waves on magnetic fluids would deserve further studies in particular theoretically, and could lead to a better understanding of plasma by studying them more easily. The analogy discovered here could be pushed forward to explore open questions in wave turbulence such as the properties of large scales~\cite{ARFM2022}, or the critical balance separating weak turbulence from strong turbulence~\cite{Meyrand2016}. The phenomenon reported here could also be observed experimentally using a ferrofluid with a high magnetic susceptibility and a high saturation magnetization. However, such ferrofluids are currently still too viscous (at least ten times the water value) which would prevent wave interactions to occur. Finally, similar effects should be expected in electrohydrodynamics of nonconducting liquids in a strong electric field due to the equivalence between the underlying equations~\cite{KochurinJETP19}.

\begin{acknowledgments}
E. Kochurin's work on developing the numerical tools is supported by Russian Science Foundation, project No. 21-71-00006. E. Falcon thanks partial support of the French National Research Agency (ANR DYSTURB project No. ANR-17-CE30-0004), and of the Simons Foundation MPS N$^{\rm o}$651463-Wave Turbulence.
\end{acknowledgments}

\end{document}